\documentclass[journal,twocolumn]{IEEEtran}
\usepackage{amsfonts}
\usepackage[tbtags]{amsmath}
\usepackage{graphicx,subfigure}
\usepackage{color}      
\usepackage{epsfig}
\usepackage{amssymb}
\usepackage{tipa}
\usepackage{bbm}

\long\def\comment#1{}

\newcommand{\beq}{\begin{equation}}
\newcommand{\eeq}{\end{equation}}
\newcommand{\beqno}{\begin{equation*}}
\newcommand{\eeqno}{\end{equation*}}

\newcommand{\bes}{\begin{split}}
\newcommand{\ees}{\end{split}}
\newcommand{\bdm}{\begin{displaymath}}
\newcommand{\edm}{\end{displaymath}}

\newtheorem{definition}{Definition}

\newcommand{\bd}{\begin{definition}}
\newcommand{\ed}{\end{definition}}

\newcommand{\bv}{\begin{vugraph}}
\newcommand{\ev}{\end{vugraph}}
\newcommand{\bi}{\begin{itemize}}
\newcommand{\ei}{\end{itemize}}
\newcommand{\ben}{\begin{enumerate}}
\newcommand{\een}{\end{enumerate}}

\newcommand{\bean}{\begin{eqnarray*} }
\newcommand{\eean}{\end{eqnarray*} }
\newcommand{\bea}{\begin{eqnarray} }
\newcommand{\eea}{\end{eqnarray} }

\newcommand{\ba}{\begin{array} }
\newcommand{\ea}{\end{array} }

\begin{document}

\title{Design and Analysis of $E^2RC$ Codes}

\author{\authorblockN{Cuizhu Shi, {\it{Student Member, IEEE}} and Aditya Ramamoorthy, {\it{Member, IEEE}}} \thanks{Manuscript received 1 Oct. 2008, revised 22 Jan. 2009. The material in this paper was presented in part at IEEE GlobeCom, New Orleans, LA, USA 2008 and will be presented in part at IEEE ICC, Dresden, Germany, 2009. Cuizhu Shi and Aditya Ramamoorthy are with the Department of Electrical and Computer Engineering, Iowa State University, USA (email: \{cshi, adityar\}@iastate.edu). This research was supported in part by NSF grants CNS-0721453 and ECCS-0802019.}}

\maketitle
\begin{abstract}

We consider the design and analysis of the efficiently-encodable rate-compatible ($E^2RC$) irregular LDPC codes proposed in previous work. In this work we introduce semi-structured $E^2RC$-like codes and protograph $E^2RC$ codes. EXIT chart based methods are developed for the design of semi-structured $E^2RC$-like codes that allow us to determine near-optimal degree distributions for the systematic part of the code while taking into account the structure of the deterministic parity part, thus resolving one of the open issues in the original construction. We develop a fast EXIT function computation method that does not rely on Monte-Carlo simulations and can be used in other scenarios as well. Our approach allows us to jointly optimize code performance across the range of rates under puncturing. We then consider protograph $E^2RC$ codes (that have a protograph representation) and propose rules for designing a family of rate-compatible punctured protographs with low thresholds. For both the semi-structured and protograph $E^2RC$ families we obtain codes whose gap to capacity is at most 0.3 dB across the range of rates when the maximum variable node degree is twenty.

\vspace{1mm}


\end{abstract}

\section{Introduction}

Low-density parity-check (LDPC) codes \cite{Gallager1963_LDPC} have found widespread acceptance in different areas due to their superior performance and low complexity decoding. In this paper, we investigate rate-compatible punctured LDPC codes that have the flexibility of operating at different code rates while having a single encoder-decoder pair.
Rate-compatible punctured codes are defined by specifying a systematic mother code that operates at the lowest code rate. The parity bits of higher rate codes in a rate-compatible code family are subsets of the parity bits of lower rate codes.
A number of papers have investigated issues around the design of good rate-compatible punctured LDPC codes.
The work of \cite{Ha2004_ratecompatiblepuncturingofLDPC} presents methods for finding optimal degree distributions for puncturing. In \cite{Ha2004_puncturingoffinitelengthLDPC} \cite{Ha2006_ratecompatiblepuncturingofLDPCwithshortblock} \cite{Yue2007_designofratecompatibleIRA}, algorithms for finding good puncturing patterns for a given mother code were proposed. There have also been attempts to design mother codes (along with puncturing patterns) with good performance under puncturing \cite{KimRM08_E2RC_journal}\cite{Yazdani2004_onconstructionofratecompatibleLDPC}\cite{aditya2006_isit}.

$E^2RC$ codes introduced in \cite{KimRM08_E2RC_journal} are linear-time encodable and have good puncturing performance across a wide range of code rates.
In this work we present systematic approaches for the design and analysis of $E^2RC$-like codes.
Let $H = [H_1 | H_2]$ denote the parity check matrix of a systematic LDPC code where $H_1$ denotes the systematic part and $H_2$ the parity part. We address the design of two types of codes in our work as explained below.

\begin{figure}[t]
\begin{center}
\includegraphics[scale=0.53]{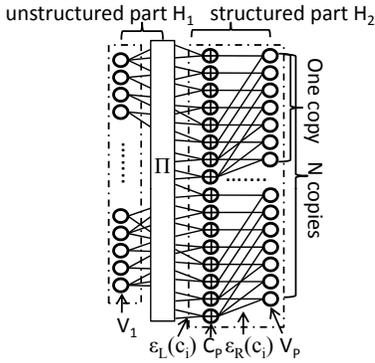}\\
\vspace{-1mm}
\caption{Tanner graph representation of $E^2RC$ codes.}
\label{figure1}
\end{center}
\end{figure}

\begin{itemize}
\item[i)] {\it Semi-structured $E^2RC$-like codes.} In these codes the parity part $H_2$ is deterministic. We use the lower triangular form introduced in \cite{KimRM08_E2RC_journal} and introduce a protograph structure for the $H_2$ part. An example is shown in Fig. \ref{figure1}. We assume a random edge interleaver between systematic variable nodes and check nodes, which divides the code into a structured part and an unstructured part, as shown in Fig. \ref{figure1}. We solve the problem of finding optimal degree distributions for the unstructured part in this case for optimizing the rate-compatible codes at any specified punctured code rate(s).
\item[ii)] {\it Structured $E^2RC$-like codes.} These codes are protograph codes as introduced in \cite{Thorpe2003_protograph}. The distinguishing feature is that the parity part of the protograph has an $E^2RC$ structure. We demonstrate that very good rate-compatible punctured code families can be obtained using the design rules we propose for the protograph construction. The protograph structure is especially valuable in practical applications as it allows parallelized decoding and requires significantly less storage space for the description of the parity-check matrix than unstructured codes when circulant permutations are used.
\end{itemize}
We obtain semi-structured $E^2RC$ codes that have a small gap to capacity across the range of puncturing rates.
Furthermore, we present optimized quasi-cyclic protograph codes based on the $E^2RC$ structure and demonstrate that very good performance can be obtained with them.

This paper is organized as follows. In Section \ref{maincontribution}, we briefly discuss the main contributions of our work.
Section \ref{ourmethod} presents our new method for the design of semi-structured $E^2RC$ codes. We also discuss the method of predicting the puncturing performance of semi-structured $E^2RC$ codes and the joint optimization of our codes at any specified punctured code rates. We explain the construction of protograph $E^2RC$ codes in Section \ref{protoconstruction}, and Section \ref{conclusion} outlines our conclusions.

\section{Main Contributions\label{maincontribution}}

\noindent
We first outline the issues left unresolved in the work of \cite{KimRM08_E2RC_journal}.
\begin{list}{}{\leftmargin=0.0cm \labelwidth=0cm \labelsep = 0cm}
\item[a)~] The original construction of $E^2RC$ codes proposed the special $H_2$ (parity part) structure of the parity-check matrix $H$. However the design of appropriate degree sequences for the $H_1$ (information part) based on the constrained $H_2$ structure, was not discussed. In \cite{KimRM08_E2RC_journal}, the authors used degree sequences designed for standard irregular codes and constructed $H$ to match these distributions as closely as possible.
\item[b)~] The construction technique did not provide any means of optimizing code performance at any particular puncturing rate or across all rates simultaneously.
\item[c)~] As pointed out by an anonymous reviewer, the original $E^2RC$ codes suffer from high error floors \cite{Ha2008_lineartimeencodable} at the mother code rate. As shown in \cite{Ha2008_lineartimeencodable}, this is because the $H_2$ structure causes the maximum check node degree to be large.
\item[d)~] The original $E^2RC$ codes work with completely random interleavers, that are hard to implement in practice.
\end{list}

In this paper, we resolve each of the issues discussed above. We briefly overview the main contributions below.
\begin{list}{}{\leftmargin=0.0cm \labelwidth=0cm \labelsep = 0cm}
\item[i)~] {\it Systematic design techniques for $E^2RC$-like codes.}\\
Note that the analysis of $E^2RC$ codes does not follow directly from the analysis of related codes such as systematic IRA codes \cite{Jin2000_IRACodes}\cite{roumya2004_designmethodforira}. This is because the structured part of IRA codes is symmetric while that of $E^2RC$ codes is quite asymmetric. In \cite{roumya2004_designmethodforira}, four methods were proposed for the design of IRA codes. The first two methods implicitly assumed one edge type in the accumulator part which was justified by the symmetry of the part. Together with a one-parameter approximation of the message distribution function, Gaussian or BEC approximation, these two methods yielded almost closed-form equations of density evolution. However, one-edge type assumption turns out not accurate enough for the structured part of $E^2RC$ codes because of its asymmetry. In the latter two methods in \cite{roumya2004_designmethodforira}, Monte Carlo simulations were used for generating the EXIT function of the structured part of IRA codes. The Monte Carlo simulation based method is accurate for computing EXIT functions of both symmetric and asymmetric constituent code components by taking the structure of the code component into account. When we design semi-structured $E^2RC$ codes using EXIT chart, we take into account the complete structure of the deterministic part of $E^2RC$ codes to compute the EXIT function as presented in Section \ref{ourmethod}. Instead of resorting to Monte Carlo simulations, we propose a fast and analytical method for computing EXIT functions by solving a set of equations. We use multiple edge types \cite{Richardson2002_multiedgeLDPC} for the structured part of $E^2RC$ codes, one edge type for each edge in the protograph representation. So instead of having only three equations (equations (19) (20) (21) in \cite{roumya2004_designmethodforira}) from the structured part of IRA codes, we have $2|E_R|+|E_L|$ equations from the structured part of $E^2RC$ codes for density evolution.
As demonstrated by simulations and the threshold predictions, this introduces a systematic method towards the design of semi-structured $E^2RC$ codes with better performance than the original $E^2RC$ codes.
\item[ii)~] {\it A fast technique for EXIT function computation of code components based on protographs.}\\
Note that usually EXIT functions are computed via Monte-Carlo simulation, which tends to be time-consuming. In this work we present a general technique for computing EXIT functions of code components with a protograph structure. This greatly speeds up the code design process. While we applied it to the design of our semi-structured $E^2RC$-like codes, it can be applied for any protograph like components, e.g. we can apply it to find the EXIT function of the $H_2$ part of the IRA code by working with its protograph representation.
\item[iii)~] {\it Simultaneous optimization of code performance across multiple rates.}\\
By exploring the $E^2RC$ structure and its designed puncturing pattern, we propose the design of good rate-compatible punctured codes so that the gap to capacity across the entire range of rates can be controlled. To the best of our knowledge, the current literature does not address this point.
\item[iv)~] {\it Alleviating the high error floor problem of the original $E^2RC$ codes.}\\
In our design of semi-structured $E^2RC$ codes, we impose a protograph structure on the $H_2$ part, which corresponds to the $H_2$ part of a very short original $E^2RC$ code. This ensures that the maximum check node degree remains low, thus preventing the high error floors that occur in the original $E^2RC$ codes at mother code rate. For a related approach see \cite{Ha2008_lineartimeencodable}.
\item[v)~] {\it Design of high-performance codes based on protographs.}\\
Codes with completely random interleavers are too complex from the point of view of implementation in hardware. In this work, we design protograph $E^2RC$ codes where both the $H_1$ and the $H_2$ parts have a protograph structure. We propose design rules for generating a family of rate-compatible protographs with good threshold properties at all punctured rates. Finally, we demonstrate codes with performance better than the original $E^2RC$ codes, that are obtained by replacing the protograph edges by circulant permutations.
\end{list}

\section{Background and Related Work\label{relatedwork}}

An LDPC code can be defined by a parity-check matrix or equivalently by a bipartite (or Tanner) graph representation.
For the bipartite graph representation, we follow the convention that a blank circle represents an unpunctured variable node participating in the transmission and a filled circle represents a punctured variable node not participating in the transmission. The asymptotic threshold of LDPC codes can be found by performing density evolution \cite{Richardson2001_designofcapacityapprochingirregularLDPC} \cite{Richardson2001_capacityofLDPCunderMPD} \cite{luby97practical} \cite{Shokrollahi1999_newsequencesoflineartimeerasurecodesapproachcapacity} on the degree distribution pair. However, for LDPC codes with structured components such as IRA codes and protograph LDPC codes \cite{Thorpe2003_protograph}, the density evolution analysis needs to take the underlying structure into account. This can be handled by classifying edges into different types \cite{Richardson2002_multiedgeLDPC} and also by using EXIT charts \cite{tenbrink2003_EXIT_IRA}. Protograph LDPC codes start with a small mini-graph (called a protograph) and construct the LDPC codes by replacing each edge in the protograph by a random permutation of a fixed size. They can be considered as a subclass of the multi-edge type LDPC codes \cite{Richardson2002_multiedgeLDPC}. Fast density evolution based on the reciprocal channel approximation \cite{Chung2000PHDthesis_ontheconstructionofsomecapacityapproachingcoding} can be performed on protographs to determine their asymptotic threshold.

\subsection{Efficiently Encodable Rate-Compatible LDPC Codes\label{e2rcoverview}}

\begin{figure}[htbp]
\begin{center}
\includegraphics[scale=0.8]{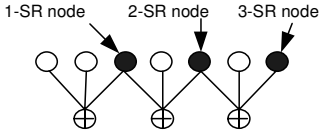}\\
\vspace{-2mm}
\caption{The figure shows an example of a 1-SR, 2-SR and 3-SR node.}
\label{figure333}
\end{center}
\end{figure}

We now briefly overview the $E^2RC$ codes introduced in \cite{KimRM08_E2RC_journal}. Let $H = [H_1 |H_2]$ denote the parity-check matrix of an $E^2RC$ code in systematic form. We say that a parity node in $H_2$ is $k$-step recoverable (or k-SR) if it can be recovered in exactly $k$ iterations of iterative decoding assuming that all the parity bits are punctured and all the systematic bits are known (Fig. \ref{figure333} shows an example). Intuitively, a large number of low-SR nodes tend to reduce the required number of decoding iterations in the high SNR regime and result in good puncturing performance.

In \cite{KimRM08_E2RC_journal}, the submatrix $H_2$ consists of exclusively degree-2 and degree-1 nodes. Moreover, when the number of parity nodes is a power of two, half the nodes in $H_2$ are 1-SR, one-fourth are 2-SR and so on. The special structure of $H_2$ for $E^2RC$ codes allows linear-time encoding and results in good puncturing performance with a puncturing pattern, where 1-SR nodes should be punctured first, 2-SR nodes be punctured next and so on depending upon the rate requirement.

The $E^2RC$ codes have good puncturing performance at relatively short block lengths. However, when the block length gets large, the structure of $H_2$ may induce a large spread in the check node degree distribution that may cause a loss of performance. In recent work, Song et al. \cite{Ha2008_lineartimeencodable} showed that $E^2RC$ codes exhibit high error floors at their mother code rate and claimed that this stems from their dispersive right degree distribution and high maximum right degree. They presented a modified approach that fixes the high error floor problem.
In Section \ref{codedesignexam}, we show that our approach also effectively eliminates the high error floors of $E^2RC$ codes at the mother code rate. In fact we obtain codes whose performance is slightly better than those in \cite{Ha2008_lineartimeencodable}.

\subsection{EXIT Chart Overview\label{exit}}

EXIT charts \cite{tenbrink2001_EXIT} were first proposed for understanding the convergence behavior of iteratively decoded parallel concatenated codes, and were later generalized to the analysis of LDPC codes \cite{tenbrink2003_EXIT_IRA} \cite{tenbrink2004_EXIT_LDPC} \cite{Alexei2004_EXIT_amodelandproperties} \cite{Sharon2006_analysisLDPCusingEXIT}. The components of an EXIT chart are the EXIT functions of the constituent code components of the iterative decoder, which relates the a priori mutual information available to a code component, denoted $I_A$ and the extrinsic mutual information generated after decoding, denoted $I_E$. The advantage of EXIT charts is that the code design problem can be reduced to a curve fitting problem between the code components (usually two in number).

For log-domain belief propagation decoding of unstructured LDPC codes, if the incoming messages to a variable node $v$ of degree $d_v$ are assumed to be Gaussian and independent, the EXIT function for the code component involving all variable nodes is given by \cite{tenbrink2004_EXIT_LDPC}
\[
I_{E,V}(I_{A,V},\sigma_{mch}^2)\hspace{60mm}
\]
\vspace{-6mm}
\begin{align}
\hspace{4mm}=\sum_{d_v}\lambda_{d_v}J(\sqrt{(d_v-1)[J^{-1}(I_{A,V})]^2+\sigma_{mch,v}^2})
\label{variable_exit}
\end{align}
where $\sigma_{mch,v}^2=4/\sigma_{n}^2$ for unpunctured $v$ ($\sigma^2_n$ represents the channel noise variance), $\sigma_{mch,v}^2=0$ for punctured $v$ and $\{\lambda_{d_v}\}$ is the edge perspective degree distribution of variable nodes.
Similarly the EXIT function for the code component involving all check nodes is given by
\[
I_{E,C}(I_{A,C})\hspace{68mm}
\]
\vspace{-6mm}
\begin{align}
\hspace{3mm}=1-\sum_{d_c}\rho_{d_c}J(\sqrt{(d_c-1)[J^{-1}(1-I_{A,C})]^2})
\label{check_exit}
\end{align}
where $\{\rho_{d_c}\}$ is the edge perspective degree distribution of check nodes.

\section{Semi-Structured $E^2RC$-like Code Design\label{ourmethod}}

In this section, we propose our design method for semi-structured $E^2RC$-like codes using EXIT charts. We consider the unstructured part and the structured part of $E^2RC$ codes shown in Fig. \ref{figure1} as the two constituent code components. This code division for EXIT chart analysis is justified by the random edge interleaver between the two code components.

We denote the set of variable nodes in the Tanner graph by $V = V_1 \cup V_2$ where $V_1$ is the subset of nodes in $H_1$ and $V_2$ is the subset of nodes in $H_2$. The check node set is denoted by $C$.
In the semi-structured $E^2RC$ codes, $H_2$ has a base protograph structure of the form proposed in \cite{KimRM08_E2RC_journal}. The base protograph shall be parameterized by the number of check nodes in it, denoted by $M$. The $H_2$ part of semi-structured $E^2RC$ codes is obtained by simply replicating the base protograph an appropriate number of times. For example, the case of $M=8$ is shown in Fig. \ref{figure1}. Check nodes are connected to the set $V_1$ by a random interleaver (denoted $\Pi$ in Fig. \ref{figure1}). We shall frequently need to refer to the protograph representation of $H_2$.
Let $V_p$ and $C_p$ denote the variable node set and the check node set in the protograph representation of $H_2$.
Let $\varepsilon(v_i)$ and $\varepsilon(c_i)$ denote the set of edges connected to $v_i \in V_p$ and $c_i \in C_p$ respectively. We shall use $\varepsilon_L(c_i)$ to denote the set of edges connecting $c_i$ and the random edge interleaver and use $\varepsilon_R(c_i)$ to denote the set of edges connecting $c_i$ and $V_p$, i.e., $\varepsilon(c_i) = \varepsilon_L(c_i) \cup \varepsilon_R(c_i)$.
Given a protograph structure on $H_2$, the problem of code design becomes one of finding good degree distributions for the variable nodes in $V_1$ and that for the edges in $\cup_{c_i \in C_p} \varepsilon_L(c_i)$ (henceforth referred to as the {\it{left check degree distribution}}). In our examples, we only consider concentrated or near-concentrated total check degrees. We have found experimentally that these tend to give the best performance. Note that since the $H_2$ part is fixed, this implies that the left check degree distribution is also more or less fixed. Accordingly in our design process we experiment with a few check degree distributions and focus on optimizing the degree distribution for the nodes in $V_1$.

We explain our design method in the context of the binary-input AWGN (BIAWGN) channel. It can be adapted to the BEC and other channels in a straightforward manner. Suppose that we are given a channel noise variance $\sigma_{n}^2$, the protograph specifying $H_2$ and the left check degree distribution.
The code design problem is to find the degree distribution $\{\lambda_{d_v}, v\in V_1\}$ so as to minimize the gap between code rate $R$ and channel capacity $C$ (corresponding to $\sigma_n^2$), while constraining the maximum variable node degree to be $d_{v, \max}$. Denote the EXIT function of the structured part by $I_{E,S}(I_{A,S})$. For a given $\{\lambda_{d_v}, v\in V_1\}$, the EXIT function of the unstructured part (see Fig. \ref{figure1}) can be expressed as (according to (\ref{variable_exit}))
\[
I_{E,unS}(I_{A,unS},\sigma_{mch}^2)\hspace{55mm}
\]
\vspace{-5mm}
\begin{eqnarray}
\hspace{4mm}=\sum_{d_v}\lambda_{d_v}J(\sqrt{(d_v-1)[J^{-1}(I_{A,unS})]^2+\sigma_{mch}^2})
\label{exituns}
\end{eqnarray}

\noindent
The code design or optimization problem is formulated as
\begin{align*}
\text{minimize~:} & ~ C - R \\
\text{subject to~:} & ~ 1.~ \sum_{d_v=1}^{d_{v,max}}\lambda_{d_v}=1, \lambda_{d_v}\geq 0\\
& ~ 2.~ ~I_{E, unS}(I_{A,unS}) > I_{A, S}(I_{E,S})\\
& ~ ~ ~ ~\mbox{ for } I_{A,unS}=I_{E,S}\in[0,1)
\end{align*}
Here, the second constraint is the zero-error constraint by ensuring the tunnel between the two EXIT curves. It is easy to see that minimizing $C - R$ for a fixed $\sigma^2_{n}$ is equivalent to maximizing $\sum_{d_v=1}^{d_{v,max}}\frac{\lambda_{d_v}}{d_v}$ for $v\in V_1$.
The computation of $I_{E,S}(I_{A,S})$ will be elaborated on in Section \ref{newmethod}. $I_{E, unS}(I_{A,unS})$ is a linear function of $\{\lambda_{d_v}, v\in V_1\}$ as in (\ref{exituns}). Therefore by a fine enough discretization of the interval $[0,1)$, we can express the above optimization as a linear program.

In practice, to set up the second constraint, we need to find the inverse map $I_{A,S}(I_{E,S})$ by using linear interpolation. We have found that a large number (say $10^4$) of ($I_{A,S},I_{E,S}$) pairs for the function $I_{E,S}(I_{A,S})$ are necessary to ensure the accuracy of the inverse map $I_{A,S}(I_{E,S})$ and the solution to the optimization problem.
By solving the above optimization problem at a certain channel parameter $\sigma^2_{n}$, we get a code of rate corresponding to the $\{\lambda_{d_v}\}$ returned from the optimization. To get an optimized code at rate $R_o$, we need to solve the above optimization problem at closely spaced channel parameter levels below the Shannon limit corresponding to $R_o$ until we get a code rate close enough to $R_o$. This necessitates numerous computations of $I_{E,S}(I_{A,S})$ and motivates the need for a fast method for computing $I_{E,S}(I_{A,S})$.

\subsection{A New Method for Computing EXIT Function of the Structured Part\label{newmethod}}

The usual approach for finding the EXIT function of a constituent code component is proposed in \cite{tenbrink2001_EXIT} by using Monte Carlo simulations. A large number of Monte Carlo simulations are needed for obtaining smooth EXIT functions. Moreover, this needs to be repeated at many different channel parameters. This makes the process rather time-consuming.

Here, we present a fast and accurate method for computing EXIT functions of structured code components of LDPC codes, such as the structured part of $E^2RC$ codes and that of IRA codes, without resorting to Monte Carlo simulations.

For convenience, we use the notation $E_R=\cup_{v_i \in V_p}\varepsilon(v_i)$ and $E_L=\cup_{c_i \in C_p}\varepsilon_L(c_i)$. Note that $\cup_{v_i \in V_p}\varepsilon(v_i)=\cup_{c_i \in C_p}\varepsilon_R(c_i)$. Suppose that the a priori inputs carried on $e \in E_L$ have average mutual information $I_{A,in}$ and that $v \in V_p$ has channel inputs parameterized by $\sigma^2_{mch,v}$. We are interested in finding $I_E$, the average mutual information associated with the extrinsic outputs carried on $e \in E_L$ after iterative decoding.
For an edge $e$ connected to node $v_i (c_i)$, we shall use the notation $I^{v_i}_{A, e} (\text{likewise~} I^{c_i}_{A, e})$ to denote the mutual information describing the a priori inputs on it and $I^{v_i}_{E,e} (\text{likewise~}I^{c_i}_{E,e})$ the mutual information describing the extrinsic outputs on it. We set up the following system of equations for the given structured code component. For $e \in E_R$ and $v_i\in V_p$, we have
\begin{align}
\label{eq:var_node_update}
I_{E, e}^{v_i} = J\bigg{(}\sqrt{ \sum_{e' \in \varepsilon(v_i) \backslash \{e\}} [J^{-1}(I_{A, e'}^{v_i})]^2 + \sigma^2_{mch, v_i} } \bigg{)}.\hspace{7mm}
\end{align}
Similarly, for $e \in E_R$ and $c_i\in C_p$,
\[
I_{E, e}^{c_i} = 1 - J\bigg{(}\bigg{(}\sum_{e' \in \varepsilon_L(c_i)} [J^{-1}(1 - I_{A, in})]^2\hspace{21mm}
\]
\vspace{-4mm}
\begin{align}
\label{eq:check_node_update_1}
\hspace{15mm}+ \sum_{e' \in \varepsilon_R(c_i) \backslash \{e\}} [J^{-1}(1 - I_{A, e'}^{c_i})]^2 \bigg{)}^{\frac{1}{2}} \bigg{)},
\end{align}
and for $e \in E_L$ and $c_i\in C_p$,
\[
I_{E, e}^{c_i} = 1 - J\bigg{(}\bigg{(}\sum_{e' \in \varepsilon_L(c_i) \backslash \{e\}} [J^{-1}(1 - I_{A, in})]^2\hspace{18mm}
\]
\vspace{-4mm}
\begin{align}
\label{eq:check_node_update_2}
\hspace{25mm}+ \sum_{e' \in \varepsilon_R(c_i)} [J^{-1}(1 - I_{A, e'}^{c_i})]^2 \bigg{)}^{\frac{1}{2}} \bigg{)}.
\end{align}
For each edge $e \in E_R$, there are two equations in the form of (\ref{eq:var_node_update}) and (\ref{eq:check_node_update_1}) respectively and two unknown variables $I_{E, e}^{v_i}$(or $I_{A, e}^{c_j}$), $I_{E, e}^{c_j}$(or $I_{A, e}^{v_i}$) associated with it; while for each edge $e \in E_L$, there is one equation in the form (\ref{eq:check_node_update_2}) and one unknown variable $I_{E, e}^{c_i}$ associated with it. We want to compute $I_{E, e}^{c_i}$ for $e \in E_L$. There are totally $2|E_R|+|E_L|$ equations and the same number of unknown variables involved in this system of equations. Note that the specific expressions for this system of equations are totally dependent on the structure of the code component. The main idea behind our method for computing the EXIT function is to find the solution to this system of equations for a given value of $I_{A, in}$ and channel parameter. We now present an intuitive method for solving this system of equations, which works in an iterative manner by applying the sequence of updates described in equations (\ref{eq:var_node_update}), (\ref{eq:check_node_update_1}) and (\ref{eq:check_node_update_2}). The details are given below.
\begin{list}{}{\leftmargin=0.0cm \labelwidth=0cm \labelsep = 0cm}
\item[1)~] {\it Problem Instance.} Given a structured code component, solve the system of equations described in (\ref{eq:var_node_update}), (\ref{eq:check_node_update_1}) and (\ref{eq:check_node_update_2}) above. The unknown variables involved in this system of equations are $I_{E, e}^{v_i}$(or $I_{A, e}^{c_j}$), $I_{E, e}^{c_j}$(or $I_{A, e}^{v_i}$) for all $e \in E_R$ and $I_{E, e}^{c_i}$ for all $e \in E_L$.
    The known variables are $I_{A, e}^{c_i} = I_{A, in}$ for all $e \in E_L$, and the channel parameter $\sigma_n^2$ from which $\sigma_{mch, v_i}^2$ can be determined for each $v_i$.
\item[2)~] {\it Initialization.} Initialize all unknown variables to be 0.
Set a small value of $\epsilon_{thresh} = 10^{-6}$.
\item[3)~] {\it Iterative Updates.}
\begin{itemize}
\item[(a)] {\it Check node update.} For all $e \in E_R$, compute $I_{E, e}^{c_i}$ using equation (\ref{eq:check_node_update_1}). Check to see whether the norm of the difference between this newly computed set of $I_{E, e}^{c_i}$ and the previously computed ones is smaller than $\epsilon_{thresh}$. If yes, then terminate; otherwise, set $I_{A, e}^{v_j} = I_{E, e}^{c_i}$ if $v_j$ and $c_i$ are connected by $e$.
\item[(b)] {\it Variable node update.} For all $e \in E_R$, compute $I_{E,e}^{v_i}$ using equation (\ref{eq:var_node_update}). Set $I_{A, e}^{c_j} = I_{E,e}^{v_i}$ if $c_j$ and $v_i$ are connected by $e$. Go to step 3(a).
\end{itemize}
\item[4)~] {\it Compute $I_{E,e}^{c_i}$ for $e\in E_L$.} For all $e \in E_L$, compute $I_{E,e}^{c_i}$ using equation (\ref{eq:check_node_update_2}). The average of these $I_{E,e}^{c_i}$ is denoted by $I_E$ and ($I_{A,in},I_E$) is a point on the EXIT function.
\end{list}
The method can be adapted for computing EXIT functions over other channels by using appropriate update equations. Moreover, it can be used to compute the EXIT function of the structured part of other codes that have a succinct protograph representation such as IRA codes.

We demonstrate the effectiveness of our method by comparing EXIT functions computed by our method and by the Monte Carlo simulation based method for two cases: the structured part of $E^2RC$ codes and that of IRA codes on BIAWGN channels respectively.
The structured part of the $E^2RC$ code has a protograph structure of size 128. All check nodes have degree 8. To get smooth curves, we apply $10^6$ a priori inputs in Monte Carlo simulations for computing each point on the curves. As shown in Table \ref{table0}, the maximum absolute error (MAE) between the EXIT functions computed using the two methods is less than $0.0072$.

\begin{table}[htbp]
\caption{Comparison of approaches for computing EXIT functions}
\vspace{-3mm}
\begin{center}
\resizebox{0.9\columnwidth}{!}{
\begin{tabular}{|l|l|l|l|l|}
\hline
\hline
\multicolumn{5}{|l|}{AWGN: noise variance$=0.95775$;$10^4~ (I_A, I_E)$ pairs generated}\\
\hline
&\multicolumn{2}{c|}{$E^2RC$}&\multicolumn{2}{c|}{IRA}\\
\hline
Method&Proposed&Simulation&Proposed&Simulation\\
\hline
Computing time (s)&3.7&24596&0.6&175886\\
\hline
MAE&0.0072&-&0.00719&-\\
\hline
\end{tabular}
}
\end{center}
\label{table0}
\vspace{-5mm}
\end{table}

\subsection{Code Design Examples\label{codedesignexam}}

\begin{figure}[htbp]
\centering
\subfigure{
\includegraphics[scale=0.55]{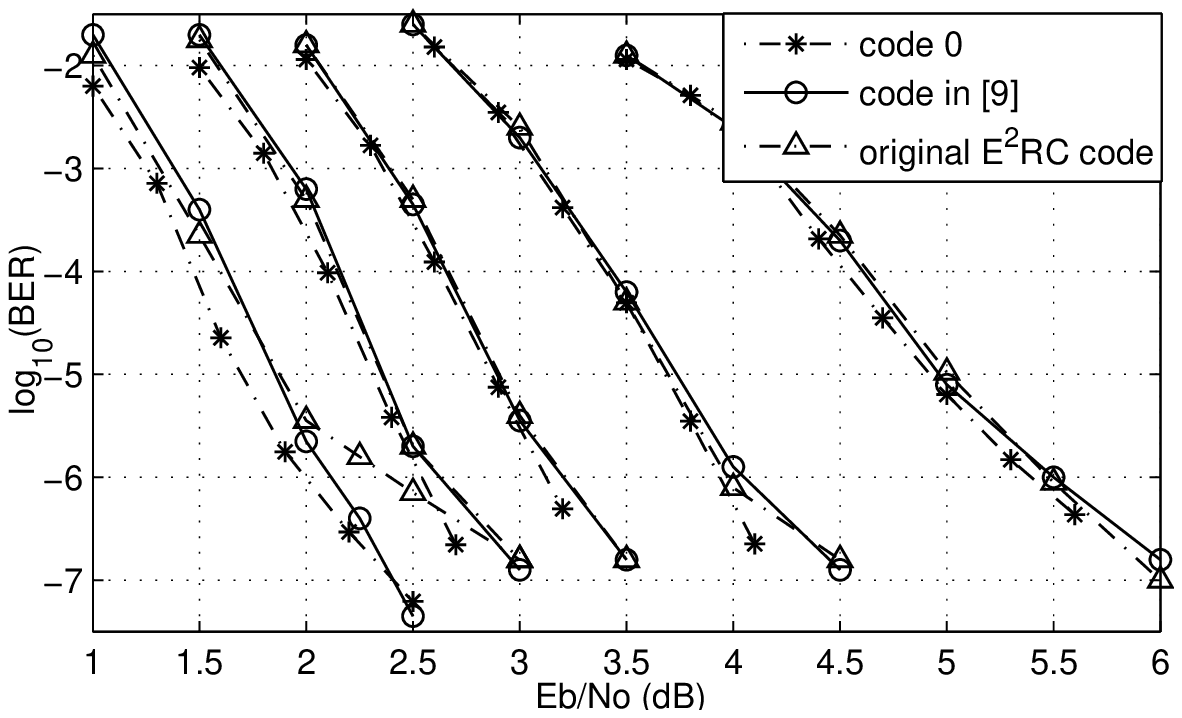}
\label{fig:subfig1}
}
\subfigure{
\includegraphics[scale=0.55]{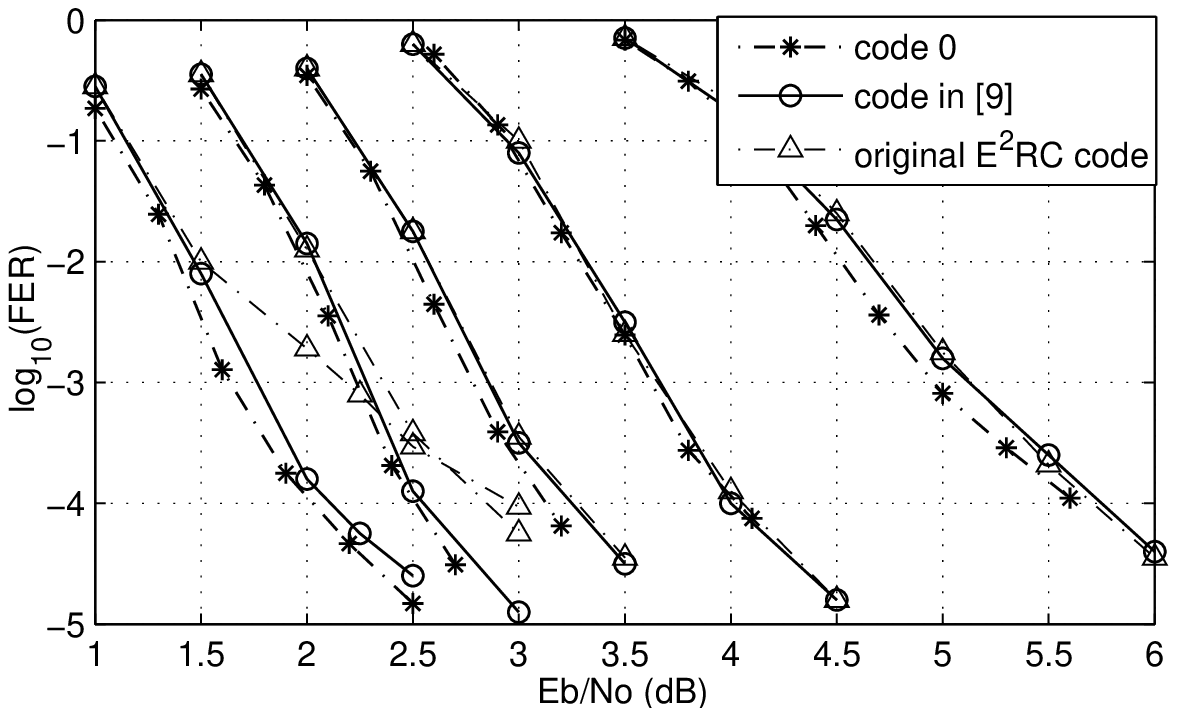}
\label{fig:subfig2}
}
\vspace{-2mm}
\caption[Code design examples]{Comparison between our code $0$ and the reference code in \cite{Ha2008_lineartimeencodable} of block length $2048$ bits. The code rates are $0.5, 0.6, 0.7, 0.8$ and $0.9$ from left to right. The figure on the top (bottom) corresponds to BER (FER).}
\label{figure4}
\end{figure}

In our first example, we design a semi-structured $E^2RC$ code with $d_{v,\max}=7$ and check degree distribution of $\rho_6=0.339623, \rho_7=0.660377$. Thus, from a complexity perspective these codes are comparable to the first design example in Section V in \cite{Ha2008_lineartimeencodable}. The mother code is of rate $0.5$ and the $H_2$ part has a protograph structure of size $M = 32$. Our optimized code (referred to as code $0$) is specified by $\lambda_3=0.4243,\lambda_7=0.5757$ for $v\in V_1$ and has an asymptotic gap of $0.38$ dB to capacity at rate $0.5$. Fig. \ref{figure4} gives the simulation results of this code of block length $2048$ bits generated by the algorithms in \cite{Tao2004_ACE}\cite{RamamoorthyW04_shortblocklengthLDPC}. For comparison, we also list the simulation results of the reference code and original $E^2RC$ code from \cite{Ha2008_lineartimeencodable}. In this paper, our codes follow the designed puncturing patterns of original $E^2RC$ codes in \cite{KimRM08_E2RC_journal} to get all puncturing code rates. From the simulation results it is clear that our code is better than the codes in \cite{Ha2008_lineartimeencodable} for all code rates. In particular they do not suffer from the high error floor problem of original $E^2RC$ codes at the mother code rate.

In our second example, we design another semi-structured $E^2RC$ codes with concentrated check degree $8$ and $d_{v,max}=20$. The optimized code (referred to as code $1$) is given by $\lambda_3=0.305825,\lambda_7=0.213474,\lambda_8=0.181737,\lambda_{20}=0.298964$ for $v\in V_1$ and it has an asymptotic gap of $0.217$ dB to capacity at rate $0.5$ which is smaller than code $0$. This is expected since $d_{v,max}$ is higher in this case. The simulation results of code $1$ of block length $16384$ bits are given in Fig. \ref{figure6}. Also given are the simulation results of the $E^2RC$ code that is constructed according to the degree distribution specified in \cite{KimRM08_E2RC_journal} (referred to as original $E^2RC$). It shows that our code achieves slightly better performance at rates near mother code rate but suffers a little at higher code rates.

We use the following terminology in this paper. The predicted threshold refers to the decoding threshold from asymptotic code performance analysis and the measured threshold refers to the channel parameter where the code achieves BER $=10^{-4}$ in simulations. For our code $0$ of length $2048$ bits, the measured threshold at rate $0.5$ is $1.47$ dB which is $0.9$ dB away from the predicted one. For our code $1$ of length $16384$ bits, this gap is only $0.4$ dB.

\subsection{Puncturing Performance Analysis and Joint Optimization of Semi-Structured $E^2RC$ Codes\label{puncturingperformance}}

The puncturing performance of a given code is specified in terms of its decoding thresholds at all punctured code rates. The given semi-structured $E^2RC$ codes are specified by $\lambda_{d_v},v\in V_1$ and the knowledge of the protograph structure of the structured part. For a given channel parameter, we can compute the two EXIT functions using (\ref{exituns}) and the approach of Section \ref{newmethod}. Note that when computing EXIT functions at different puncturing code rates, we follow the designed puncturing pattern of $E^2RC$ codes. The decoding threshold at a given code rate is determined by finding the channel parameter where the two EXIT curves (computed under the puncturing pattern at that rate) just begin to separate.

\begin{figure}[htbp]
\centering
\includegraphics[scale=0.65]{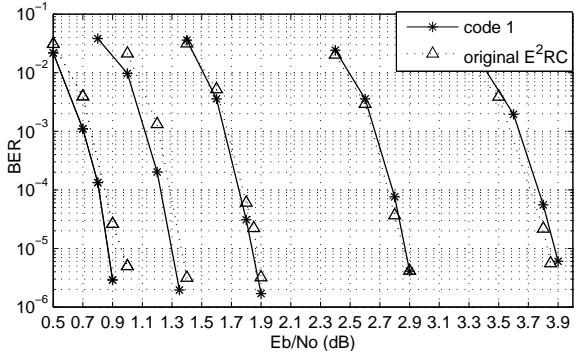}
\vspace{-2mm}
\caption[Code design examples]{Comparison between our second code example and original $E^2RC$ code in \cite{KimRM08_E2RC_journal} of block length $16384$ bits. The code rates are $0.5, 0.5714, 0.6667, 0.8$ and $0.8889$ from left to right. }
\label{figure6}
\end{figure}

Our puncturing performance analysis of code $1$ suggests that it has asymptotic decoding thresholds of around $0.40$, $0.85$, $1.40$, $2.45$, $3.44$ dB at rates $\frac{8}{16},\frac{8}{14},\frac{8}{12},\frac{8}{10}$ and $\frac{8}{9}$ respectively. The measured thresholds at these rates for the code of block length $16384$ bits based on the simulation results in Fig. \ref{figure6} are around $0.80, 1.22, 1.75, 2.78, 3.76$ dB, which are consistent with the predicted ones with gaps uniformly around $0.35$ dB.

\vspace{1mm}
\noindent \underline{{\it Joint Optimization of Semi-Structured $E^2RC$ Codes\label{joint}}}\\
We now demonstrate that we can design our codes such that they have a small gap to capacity at all puncturing rates. This is because our puncturing pattern is deterministic and allows the determination of the asymptotic threshold at any puncturing rate for any given $\{\lambda_{d_v},v\in V_1\}$. Let $\mathbf{R}$ be a specified set of code rates where we want to optimize the code. Let $\sigma(g, R_i)$ denote the channel noise parameter that is at a gap of $g$ from the channel parameter corresponding to the Shannon limit at rate $R_i$. Let $I_{A, S}(I_{E,S},\sigma(g,R_i))$ denote the plot of $I_{A,S}$ vs. $I_{E, S}$ under the puncturing pattern corresponding to rate $R_i$, at the channel parameter $\sigma(g,R_i)$. The notation $I_{A, unS}(I_{A,unS},\sigma(g,R_i))$ will be used analogously. We can formulate the joint optimization problem as minimizing the maximum gap to capacity at all rates in $\mathbf{R}$ as follows.

\vspace{1mm}
\noindent
\underline{{\bf{Joint optimization algorithm}}}

\noindent
{\bf{for}} $g=g_{min}:g_{max}$

{\bf{Solve}} the following linear program optimization problem
\begin{align*}
\mbox{maximize:}&\sum_{d_v=1}^{d_{v,max}}\frac{\lambda_{d_v}}{d_v}\\
\mbox{subject to:}&1.~\sum_{d_v=1}^{d_{v,max}}\lambda_{d_v}=1, \lambda_{d_v}\geq 0,\\
&2.~I_{E, unS}(I_{A,unS},\sigma(g,R_i)) > I_{A, S}(I_{E,S},\sigma(g,R_i))\\
&\mbox{ ~ for~} I_{A,unS}=I_{E,S}\in[0,1), \text{~for all~} R_i \in \mathbf{R}
\end{align*}

{\bf{if}} mother code rate corresponding to $\{\lambda_{d_v}\}$ is acceptable

\hspace{4mm} break; return $\{\lambda_{d_v}\}$ and $g$.

{\bf{endif}}

\noindent
{\bf{endfor}}
\vspace{1mm}

\noindent
Note that the second set of constraints is the zero-error constraint by ensuring an iterative decoding tunnel for all EXIT charts at all the required code rates. We can obtain all the required EXIT functions relatively quickly using our approach outlined previously under the puncturing patterns for each $R_i$. To obtain optimized $\{\lambda_{d_v}\}$, we basically keep increasing $g$ until we get the desired code rate. The code specified by $\lambda_{d_v}$ is guaranteed to have asymptotic performance gap to capacity no larger than $g$ at all code rates in $\mathbf{R}$.

\begin{figure}[htbp]
\begin{center}
\includegraphics[scale=0.65]{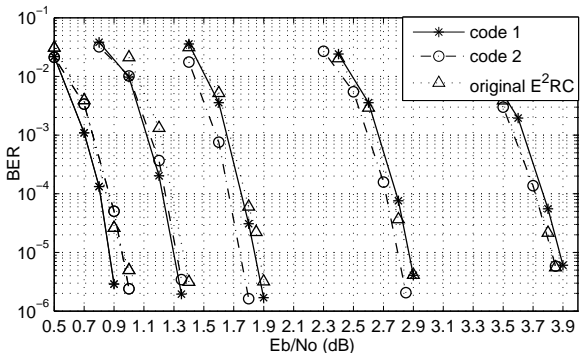}
\vspace{-2mm}
\caption{Comparison between two semi-structured $E^2RC$ codes optimized at mother code rate (code $1$) and simultaneously optimized at multiple rates (code $2$) and original $E^2RC$ code of block length $16384$ bits. The code rates are $0.5, 0.5714, 0.6667, 0.8$ and $0.8889$ from left to right.}
\label{figure7}
\end{center}
\end{figure}

We designed a semi-structured $E^2RC$ code that was jointly optimized across the rate range $\frac{8}{16} \sim \frac{8}{9}$, where $M=32$, all check nodes have degree $8$ and $d_{v,\max}=20$. The code is specified by $\lambda_3=0.309090, \lambda_6=0.278794, \lambda_{20}=0.412116$.
Fig. \ref{figure7} gives the simulation results for the code of block length $16384$ bits (listed as code $2$). Also plotted are the simulation results in Fig. \ref{figure6} for code $1$ and original $E^2RC$ code from Section \ref{codedesignexam}. The puncturing performance analysis suggests that code $1$ has asymptotic performance gaps around $0.21, 0.32, 0.34, 0.41, 0.405$ dB to capacity at rates $\frac{8}{16},\frac{8}{14},\frac{8}{12},\frac{8}{10}$ and $\frac{8}{9}$ respectively while code $2$ has much more uniform gaps of around $0.29, 0.30, 0.25, 0.29, 0.295$ dB to capacity at these rates. The simulation results in Fig. \ref{figure7} also suggest uniformly better performance of code $2$ compared to code $1$ across the range of rates. Moreover, the measured thresholds of the two codes at all rates in simulations are in good agreement with the predicted thresholds.
At all code rates, the gap between the measured threshold and the predicted threshold is around $0.35$ dB. Finally, we note that code $2$ achieves better or at least the same performance as original $E^2RC$ code at all rates.

The methods described in this section apply more generally to codes that have a structured component with a protograph representation, such as IRA codes. We applied our method to the design of IRA codes as well and obtained jointly optimized codes with performance gaps around $0.27, 0.30, 0.195, 0.27, 0.29$ dB to capacity at rates $\frac{8}{16}$ ,$\frac{8}{14}$, $\frac{8}{12}$, $\frac{8}{10}$ and $\frac{8}{9}$ respectively. Though not presented here, the simulation results of the IRA code are almost identical to the jointly optimized $E^2RC$ code (code $2$) discussed above.

\begin{figure}[htbp]
\begin{center}
\includegraphics[scale=0.53]{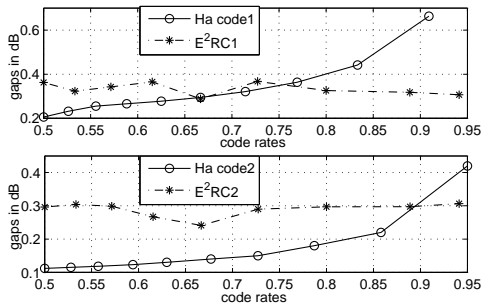}
\vspace{-2mm}
\caption{Asymptotic performance comparison between our codes and those from \cite{Ha2004_ratecompatiblepuncturingofLDPC}.}
\label{figure9}
\end{center}
\end{figure}

To the best of our knowledge, a joint optimization algorithm that minimizes the gap to capacity simultaneously across all code rates has not been considered previously in the literature.
In \cite{Ha2004_ratecompatiblepuncturingofLDPC}, the authors found the optimal puncturing patterns for two optimized mother codes (referred to as Ha code 1 and Ha code 2) and also gave their asymptotic puncturing performance using Gaussian approximation based density evolution. In Fig. \ref{figure9}, we compare the asymptotic thresholds of their codes with our code families. Note that our codes have the same maximum variable node degree and slightly smaller average variable node degree compared to the codes in \cite{Ha2004_ratecompatiblepuncturingofLDPC}. We observe that the gaps to capacity for our codes remain more or less the same across the range of rates, whereas the codes in \cite{Ha2004_ratecompatiblepuncturingofLDPC} exhibit a larger gap to capacity at higher code rates.

\section{Protograph $E^2RC$ codes construction\label{protoconstruction}}

In this section, we introduce the construction of a class of structured $E^2RC$-like codes based on protographs. The basic idea in these codes is to impose a protograph structure on the systematic part $H_1$ of the parity-check matrix as well (in addition to the protograph structure on $H_2$). We obtain a family of protographs with asymptotic gaps to capacity no larger than $0.28$ dB across a wide range of rates when the maximum variable node degree is twenty. These codes have excellent finite length performance as well.
In this part of the work, we use the reciprocal channel approximation of density evolution for computing the threshold for a given protograph \cite{Thorpe2003_protograph}\cite{Chung2000PHDthesis_ontheconstructionofsomecapacityapproachingcoding}\cite{Richardson2002_multiedgeLDPC}.
The construction algorithm goes as follows.
\begin{list}{}{\leftmargin=0.0cm \labelwidth=0cm \labelsep = 0cm}
\item[1)~] {\it Find a good high-rate protograph.} Using density evolution we first identify a high-rate protograph (starting protograph) with a low threshold.
\item[2)~] {\it Use check-splitting to obtain lower-rate protographs with low thresholds.} From the starting protograph, we perform check splitting in a systematic manner to obtain a family of good rate-compatible punctured protographs where the parity part has the $E^2RC$ structure. These correspond to different code rates of the rate-compatible code family.
\item[3)~] {\it Construct the LDPC code by replacing protograph edges with carefully chosen circulant permutations.} With the protograph of mother code rate constructed from above steps, a larger graph defining the LDPC code is constructed by replacing the protograph edges with appropriately chosen circulant permutations by using techniques in \cite{Hu2001_PEG}\cite{Tao2004_ACE}\cite{RamamoorthyW04_shortblocklengthLDPC}\cite{aditya2004_loweringerrorfloor}.
\end{list}
In the sequel we shall attempt to explain the construction process by means of an example. However, it should be clear that the techniques are applicable in general.

\vspace{-1mm}
\subsection{Starting Protograph\label{startingproto}}

Let the desirable code rate range for the code family be $R_{min} \leq R \leq R_{max}$. A high-rate protograph of size $M_0 \times N_0$ ($M_0$ - number of check nodes, $N_0$ - number of variable nodes) with low threshold serves as the starting protograph.
The mother code protograph is of size $M \times N$ such that $\frac{N-M}{N} \leq R_{min}, \frac{N_0-M_0}{N_0} \geq R_{max}$ and $N-M = N_0-M_0$.
These conditions guarantee that the desirable code rate range is achievable by the construction. In addition, these parameters should be kept relatively small (less than $50$) to keep the construction complexity manageable.
We impose the constraint that the degree of the variable nodes in the starting protograph is at least three to avoid a high error floor. We perform an exhaustive search using the reciprocal channel approximation of density evolution \cite{Chung2000PHDthesis_ontheconstructionofsomecapacityapproachingcoding}\cite{Richardson2002_multiedgeLDPC} to find a protograph with low threshold of size $M_0 \times N_0$ and with variable node degree between 3 and a maximum degree $d_{v,max}$.
In our example, $R_{min}=0.5$, $R_{max}=\frac{8}{9}$, $d_{v,max}=20$ and the size parameters are decided as $M_0=1, N_0=9$ and $M=8, N=16$.
Our example starting protograph consists of one check node and nine variable nodes of degree \{20,8,3,3,3,3,3,3,3\} respectively. i.e. each variable node is connected to the single check node by multiple edges. It has a threshold (computed according to \cite{Chung2000PHDthesis_ontheconstructionofsomecapacityapproachingcoding}\cite{Richardson2002_multiedgeLDPC}) of $3.27$ dB which is $0.24$ dB away from capacity.

\vspace{-1mm}
\subsection{Check-Splitting}\label{sec:check-splitting}

\begin{figure}[htbp]
\vspace{-1mm}
\begin{center}
\includegraphics[scale=0.78]{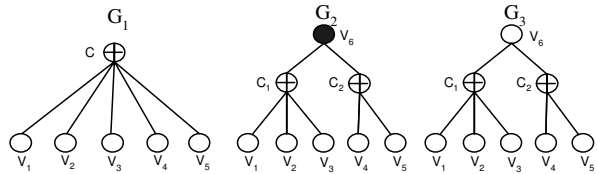}\\
\vspace{-2mm}
\caption{Check node $c$ with degree-5 is divided into $c_1$ and $c_2$. The new check nodes $c_1$ and $c_2$ are connected by a new degree-2 variable node $v_6$.}
\label{figure11}
\end{center}
\end{figure}

The operation of check-splitting (also used in \cite{Divsalar2006_protoLDPCwithlinearmindistance}) on a check node $c$ of degree $d$ in the protograph $G_1$ proceeds as shown in Fig. \ref{figure11}. We split $c$ into two new check nodes $c_1$, of degree $d_1$, and $c_2$, of degree $d_2$, such that $d = d_1 + d_2$. Next we introduce a new variable node $v_6$ and introduce edges $c_1 - v_6$ and $c_2 - v_6$, so that the degree of $v_6$ is two. The resultant protograph $G_3$ is of lower rate than $G_1$.

Starting with the high-rate starting protograph, we apply check-splitting repeatedly in a specific manner, and finally arrive at the protograph of low rate mother code. In fact, the protographs produced in the check-splitting process form a family of rate-compatible protographs if we consider the newly added degree-two variable nodes in check-splitting as parity nodes providing incremental redundancy. However, the check-splitting needs to be done carefully in order to have good code performance across all rates. We note that performing density evolution on $G_1$ in Fig. \ref{figure11} can predict the threshold of $G_3$ under puncturing (upon puncturing $v_6$, the resultant $G_2$ has the same asymptotic decoding threshold as $G_1$) \cite{Hossein2007_resultsonpuncturedLDPCandimproveddecoding}.

\subsection{Constructing Protographs with $E^2RC$-like Structure}\label{sec:construction-e2rc}

We shall call the original variable nodes of degree at least three in the starting protograph, {\it old nodes} and the variable nodes of degree two introduced in check-splitting, {\it new nodes}.
For a given check node, we define its old (new) node degree to be the number of connections to the old (new) nodes.
When we split $c_0$ into $c_{01}$ and $c_{02}$, a decision needs to be made on how the connections of $c_0$ are divided between them.
To obtain the $E^2RC$ structure, $c_{01}$ is allocated all of $c_0$'s new node degree; the parity node newly introduced in check-splitting has one connection to both $c_{01}$ and $c_{02}$ (for a proof see the Appendix). The old node degree also needs to be divided between $c_{01}$ and $c_{02}$ in a manner that ensures that the threshold of the new protograph is low. We discuss it in more detail in Section \ref{splittingpattern}.

At each stage of the construction, we perform check-splitting on all check nodes in the current protograph. We use the example starting protograph from Section \ref{startingproto} to demonstrate the process. Here $M_0=1$ and $M = 8$, so we shall have $\log_{2} 8 = 3$ construction stages.
Let $M_{n_s}$ denote the number of check nodes in the protograph at the beginning of stage $n_s$. In stage $n_s$, we perform check-splitting on all $M_{n_s}$ check nodes so that a set of $M_{n_s}$ protographs of decreasing rates are generated at this stage.
The order in which the check nodes are split can affect the thresholds at those rates.

We now show the first and second splitting stages for our starting protograph. In the first stage, check-splitting on the single check node generates a new protograph of rate $\frac{8}{10}$, shown in Table \ref{table2}.
In the second stage, there are two check nodes in the protograph. Density evolution analysis tells us that performing check-splitting on $c_{01}$ first gives a protograph of rate $\frac{8}{11}$ with a better decoding threshold.
So in this stage, we first split $c_{01}$ to generate a protograph of rate $\frac{8}{11}$
and then split $c_{02}$ to generate a protograph of rate $\frac{8}{12}$ (see Table \ref{table4}).

\begin{table}[htbp]
\caption{}
\vspace{-4mm}
\begin{center}
\resizebox{0.77\columnwidth}{!}{
\begin{tabular}{llllllllll|l}
\hline\hline
&&&&&old &&&&& new\\
\hline
         &$v_0$ & $v_1$ & $v_2$ & $v_3$ & $v_4$ & $v_5$ & $v_6$ & $v_7$ & $v_8$ & $v_9$\\
$c_{01}$ & 10 & 4 & 2 & 1 & 2 & 1 & 2 & 1 & 2 & 1\\
$c_{02}$ & 10 & 4 & 1 & 2 & 1 & 2 & 1 & 2 & 1 & 1\\
\hline
\end{tabular}
}
\end{center}
\label{table2}
\end{table}

\vspace{-6mm}

\begin{table}[htbp]
\caption{}
\vspace{-4mm}
\begin{center}
\resizebox{0.93\columnwidth}{!}{
\begin{tabular}{llllllllll|lll}
\hline\hline
&&&&&old &&&&& &new&\\
\hline
&$v_0$ & $v_1$ & $v_2$ & $v_3$ & $v_4$ & $v_5$ & $v_6$ & $v_7$ & $v_8$ & $v_9$ &$v_{10}$ & $v_{11}$\\
$c_{011}$&5&2&1&1&1&0&1&1&1&1&1&0\\
$c_{012}$&5&2&1&0&1&1&1&0&1&0&1&0\\
$c_{021}$&5&2&1&1&0&1&1&1&0&1&0&1\\
$c_{022}$&5&2&0&1&1&1&0&1&1&0&0&1\\
\hline
\end{tabular}
}
\end{center}
\label{table4}
\vspace{-1mm}
\end{table}

\begin{figure}[htbp]
\begin{center}
\includegraphics[scale=0.6]{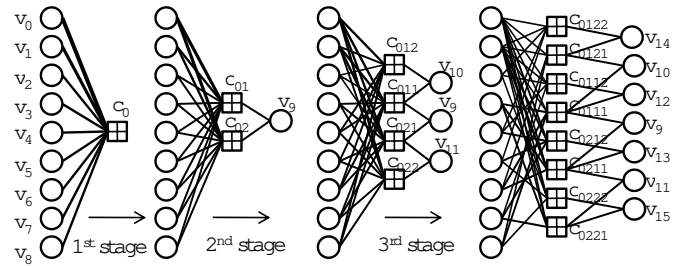}
\vspace{-2mm}
\caption{Protograph construction from check-splitting.}
\label{figure20}
\end{center}
\end{figure}

The third stage proceeds in a similar manner. Refer to Fig. \ref{figure20} for a graphical illustration of the construction process.
Note that at the end of stage $n_s$ of the algorithm, the $M_{n_s}$ newly added parity nodes are all $1$-SR. Also a k-SR node at the end of stage $n_s-1$ would become (k+1)-SR at the end of stage $n_s$ (as shown in the appendix).
In this way the construction ensures that half of the parity nodes in the final protograph are $1$-SR, one-fourth are $2$-SR and so on, i.e., our construction results in protographs with $E^2RC$ structure. The parity nodes of the resultant mother code protograph are punctured in the inverse order in which they were added in the construction in order to obtain higher puncturing rates.

\vspace{-2mm}

\subsection{Deciding Splitting Patterns\label{splittingpattern}}

In check-splitting, the connections between a given check node and the old nodes can be divided among the two new check nodes in many ways. Let $c_0$ be the check node to be split. Let $s_0$ be the vector of connections between $c_0$ and the old nodes, e.g., $s_0 = [20 ~8 ~3 ~3 ~3 ~3 ~3 ~3 ~3]$ in our example. The splitting pattern refers to the set of vectors $s_{01} = [10~4~2~1~2~1~2~1~2]$ and $s_{02} =[10~4~1~2~1~2~1~2~1]$ (shown in Table \ref{table2}) that determine the connections between $c_{01}$ and $c_{02}$ and the old nodes. Thus $s_0 = s_{01} + s_{02}$. It is easily seen that the number of possible splitting patterns are huge in each check-split and it's impossible to evaluate all of them in our construction. For example, when splitting $s_0$ into $s_{01}$ and $s_{02}$, we have totally $20^1 \times 8^1 \times 3^7=349920$ possible splitting patterns.
Our search for good splitting patterns for each check-splitting is guided by two main points.
\begin{list}{}{\leftmargin=0.0cm \labelwidth=0cm \labelsep = 0cm}
\item[a)~] {\it Trade-off between the performance of high-rate and low-rate protographs.} We have found that there exists a tradeoff between the performance of high-rate and low-rate protographs during the construction. For example, very low thresholds for the higher rate protographs typically come at the expense of higher thresholds for the low-rate protographs.

\item[b)~]{\it Equal splitting patterns give good performance.} Note that considering all possible splitting patterns at all possible stages is essentially computationally infeasible. We have found that splitting patterns that split the connections roughly equally between the two new check nodes in each check-splitting result in good thresholds across all code rates in the family. This reduces the search space a lot and it becomes possible to perform density evolution analysis to determine proper splitting patterns at each stage.
\end{list}
In Table \ref{table5}, we present one of the mother code protographs that we have constructed from the example starting protograph from Section \ref{startingproto} using the construction algorithm described above. Also shown is the gap to Shannon limit for the protograph at different puncturing rates $\frac{8}{9} \sim \frac{8}{16}$. The gap to capacity remains between 0.235-0.278 dB across the range of rates. In fact, we could choose other splitting patterns such that the protograph family has an even lower threshold at some high code rate during the early stage of the construction, but it would come at the expense of higher thresholds for the lower code rates in the later stage of the construction. These protographs are available at \cite{myhomepage} (Due to lack of space we are unable to include them here).

\begin{table}[htbp]
\caption{Mother code Protograph-1}
\vspace{-4mm}
\begin{center}
\resizebox{1\columnwidth}{!}{
\begin{tabular}{p{1mm}p{1mm}p{1mm}p{1mm}p{1mm}p{1mm}p{1mm}p{1mm}p{1mm}|p{1mm}p{1mm}p{1mm}p{1mm}p{1mm}p{1mm}p{1mm}}
\hline\hline
$v_0$&$v_1$&$v_2$&$v_3$&$v_4$&$v_5$&$v_6$&$v_7$&$v_8$&$v_9$&$v_{10}$&$v_{11}$&$v_{12}$&$v_{13}$&$v_{14}$&$v_{15}$\\
3&1&1&0&1&0&0&1&0&0&0&0&1&0&0&0\\
2&1&0&1&0&0&1&0&1&1&1&0&1&0&0&0\\
3&1&0&0&0&0&1&0&1&0&1&0&0&0&1&0\\
2&1&1&0&1&1&0&0&0&0&0&0&0&0&1&0\\
3&1&0&0&1&1&0&1&0&0&0&0&0&1&0&0\\
2&1&0&1&0&0&1&0&0&1&0&1&0&1&0&0\\
3&1&0&1&0&0&0&1&0&0&0&1&0&0&0&1\\
2&1&1&0&0&1&0&0&1&0&0&0&0&0&0&1\\
\hline
\multicolumn{16}{l}{Gap to Shannon limit in dB (rates $\frac{8}{9}\sim \frac{8}{16}$)}\\
\hline
\multicolumn{16}{l}{0.235 0.253 0.270 0.246 0.278 0.275 0.274 0.270}\\
\hline
\end{tabular}
}
\end{center}
\label{table5}
\vspace{-2mm}
\end{table}

\subsection{Results}

We compared the asymptotic thresholds of the code family represented by protograph-1 above and the AR4JA code family in \cite{Divsalar2006_protoLDPCwithlinearmindistance}. For the three common code rates $\frac{1}{2}, \frac{2}{3}$, and $\frac{4}{5}$, our protograph is better than the AR4JA family with performance gains of $0.17,0.143$ and $0.12$ dB respectively.
The average variable node degree of our codes is a little higher than that of AR4JA family. Note however that the codes in \cite{Divsalar2006_protoLDPCwithlinearmindistance} are not rate-compatible punctured codes.

\begin{figure}[htbp]
\begin{center}
\includegraphics[scale=0.65]{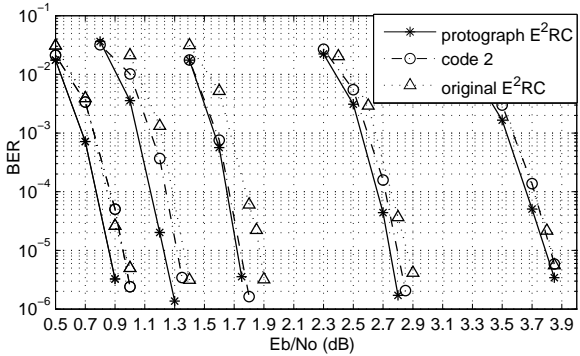}
\vspace{-2mm}
\caption{Comparison between protograph $E^2RC$ code, optimized semi-structured $E^2RC$ code and the original $E^2RC$ code of block length $16384$ bits at rates $0.5, 0.5714, 0.6667, 0.8$ and $0.8889$ from left to right.}
\label{figure13}
\end{center}
\end{figure}

We constructed protograph $E^2RC$ codes with block length $16384$ bits by replacing each edge in protograph-1 by an appropriate circulant permutation using the algorithms proposed in \cite{Tao2004_ACE}\cite{RamamoorthyW04_shortblocklengthLDPC}. Fig. \ref{figure13} gives the simulation results of this protograph $E^2RC$ code, the jointly optimized semi-structured $E^2RC$ code (code $2$) and original $E^2RC$ code of the same block length from Section \ref{joint}. From the asymptotic performance analysis for protograph $E^2RC$ code as shown on the bottom of Table \ref{table5} and for code $2$ in Section \ref{joint}, we see that protograph $E^2RC$ code is quite competitive to the optimized semi-structured $E^2RC$ code. The simulation results shown in Fig. \ref{figure13} are consistent with the analysis. Moreover, both protograph $E^2RC$ code and optimized semi-structured $E^2RC$ code achieve better performance than the original $E^2RC$ code.

\section{Conclusions\label{conclusion}}

The $E^2RC$ codes were proposed in \cite{KimRM08_E2RC_journal} as a promising class of rate-compatible codes. In this work we introduced semi-structured $E^2RC$-like codes and protograph $E^2RC$ codes. We developed EXIT chart based methods for the design of semi-structured $E^2RC$-like codes that allow us to determine near-optimal degree distributions for the systematic part of the code while taking into account the structure of the deterministic parity part. We presented a novel method for finding EXIT functions for structured code components that have a succinct protograph representation that is applicable in other scenarios as well. This allows us to analyze the puncturing performance of these codes and obtain codes that are better than the original construction. Using our approach we are able to jointly optimize the code performance across the range of rates for our rate-compatible punctured codes. Finally we consider $E^2RC$-like codes that have a protograph structure (called protograph $E^2RC$ codes) and propose design rules for rate-compatible protographs with low thresholds. These codes are useful in applications since the protograph structure facilitates implementation. For both the semi-structured and protograph $E^2RC$ families we obtain codes with small gaps to capacity across the range of rates.

\section{acknowledgement}

The authors would like to thank the anonymous reviewers and the guest editor Prof. William Ryan whose comments greatly improved the quality of the paper.

\begin{appendix}

We first recall the precise construction rule. Let $c_0$ be a check node with a certain number of connections to {\it new} nodes of degree-2. When $c_0$ is split into $c_{01}$ and $c_{02}$, $c_{01}$ inherits all of $c_0$'s connections to new nodes and both $c_{01}$ and $c_{02}$ have one connection with the newly introduced variable node.

Suppose that the starting protograph is of size $M_0 \times N_0$. At the end of stage $k$ of the construction algorithm, it is clear that there will be $M_0 \cdot 2^k$ check nodes and $M_0 (2^k  - 1)$ new degree-2 parity nodes. Our aim is to show that this construction algorithm results in an $H_2$ part with the $E^2RC$ structure. i.e. at the end of stage $k$ of the algorithm, there are $M_0 \cdot 2^{k-1}$ 1-SR nodes, $M_0 \cdot 2^{k-2}$ 2-SR nodes, \dots , and $M_0$ k-SR nodes. We proceed by induction.

{\it Base Case.} At the end of the first stage, we will have $2M_0$ check nodes and $M_0$ new degree-2, 1-SR parity nodes.

{\it Inductive Step.} Suppose that the statement is true at the end of stage $k$. We will show that it is true at the end of stage $k+1$. To see this note that at the end of stage $k+1$ we will have $M_0 \cdot 2^{k+1}$ check nodes formed by splitting the check nodes at the end of stage $k$. According to the construction algorithm, $M_0 \cdot 2^k$ check nodes will inherit the previous connections while the remaining will have just one connection to the $M_0 \cdot 2^k$ newly introduced degree-2 variable nodes. Therefore we will have at least $M_0 \cdot 2^k$ 1-SR nodes at the end of stage $k+1$. Next we note that any node that was of type $\alpha$-SR at the end of stage $k$ will now become of type ($\alpha$+1)-SR. This is because each of the check nodes it is connected to will have one additional connection. This implies that at the end of stage $k+1$ there will be $M_0 \cdot 2^{k}$ 1-SR nodes, $M_0 \cdot 2^{k-1}$ 2-SR nodes, \dots , and $M_0$ (k+1)-SR nodes. This shows the required result.
\end{appendix}

\bibliographystyle{IEEEtran}
\bibliography{DesignandAnalysisofE2RCCodes}

\end{document}